\documentclass[10pt,pra,amsmath,amssymb]{revtex4}

\usepackage{graphicx}

\newcommand{\bra}[1]{\mathop{\left\langle #1 \right|}\nolimits}
\newcommand{\ket}[1]{\mathop{\left| #1 \right\rangle}\nolimits}

\newcommand{\braket}[2]{\mathop{\left\langle #1 \right|\left. #2\right\rangle}\nolimits}

\begin{document}
\title{Quantum uniqueness}
\author{Denis Sych and Gerd Leuchs}
\affiliation{Max Planck Institute for the Science of Light, 
G\"unther--Scharowsky--Strasse 1 / Bau24,
D-91058 Erlangen, Germany}
\affiliation{Institute for Optics, Information and Photonics, University of Erlangen--N\"urnberg, Staudtstrasse 7 / B2, 91058 Erlangen, Germany}
\date{\today}
\begin{abstract}
In the classical world one can construct two identical systems which have identical behavior and give identical measurement results. We show this to be impossible in the quantum domain. We prove that after the same quantum measurement two different quantum systems cannot yield always identical results, provided the possible measurement results belong to a non orthogonal set. This is interpreted as quantum uniqueness~--- a quantum feature which has no classical analog. Its tight relation with objective randomness of quantum measurements is discussed.
\end{abstract}
\maketitle

One of the most amazing aspects of quantum theory is the prediction of phenomena which do not exist in the classical world. In the area of quantum information, the well--known examples are the special superposition states called entangled states \cite{EPR:35,Bohr:35,Horodecki:09RMP}, the violation of Bell's inequality \cite{Bell:64,Aspect:82}, teleportation \cite{Bennett:93}, the ``no--cloning'' theorem \cite{Wooters:82,Dieks:82}, etc. Initially they appeared as purely theoretical issues, but later found important applications, e.g. in cryptography and computation \cite{BEZ:00,Bruss:06,Gisin:02RMP,Scarani:09RMP}.

The ``no--cloning'' theorem refers to creating a copy of a system having the identical wavefunction as the original. In general, when the system to be copied is in an unknown state out of a non orthogonal set, it is not possible to create its perfect copy. However, under the restriction, that the system is in a fully known state or belongs to a known orthogonal set, the second system can be created in an identical state~--- a perfect clone. 

In this work, we are asking a question which goes beyond the ``no--cloning'' issue: is it possible that two systems are absolutely identical to each other with respect to their possible measurements results? Can the results of two identical quantum measurements of two quantum systems coincide universally? In other words, can a perfect clone be a perfect copy? Following everyday experience, we say that systems $A$ and $B$ are perfect copies of each other if there is no difference between them, i.e. both systems have absolutely the same properties and show absolutely the same manifestations. Particularly, identical measurements of systems $A$ and $B$ must yield identical results. Otherwise, if for a given system $A$ there is no such perfect copy $B$, then we say that the system $A$ is unique. In the classical world the behavior of systems can be described by deterministic laws. Two classical systems with the same parameters, same initial conditions and governed by the same dynamics are absolutely identical to each other. Thus the existence of a perfect copy is possible in principle, and therefore classical systems are not unique.

It is important to note, that we do not refer to the uniqueness of a quantum state itself or any other internal mathematical structure of the systems, but to the uniqueness of the system including the external characteristics which can be objectively measured. Indeed, to predict the result of a measurement of a system $A$, one needs another system $B$ (an oracle), which must give the same result as the system $A$. The system $B$ can be a real physical system, or a simulator such as a classical or a quantum computer, or even an abstract mathematical model. The only condition is that it has to follow the initial system $A$ perfectly. With respect to the measurement it means that, without loss of generality, system $B$ has to belong to an equivalent Hilbert space, and identical measurements should yield the same physical parameters as $A$. But as we demonstrate in this paper, the perfect copy of any quantum system does not exist even in principle, thus all quantum systems are unique. 

The generalized quantum measurement is described by a positive operator--valued measure (POVM) $\{\hat E(i)\}$, $\hat 1=\sum\limits_i \hat E(i)$ \cite{Helstrom:76,Kraus:83,Peres:93}. To quantify a possible discrepancy between the measurement outcomes, let us consider the divergence operator
\begin{equation}\label{divop}
\hat C_{AB}=\frac{1}{2}\sum\limits_i\left(\hat E_A(i)\otimes \hat 1_B-\hat 1_A\otimes\hat E_B(i)\right)^2.
\end{equation}
The mean value $C_{AB}={\rm Tr}[\hat\rho_{AB}\hat C_{AB}]$ of this operator reflects a possible difference of two alternatives: one measures the system $A$ and keeps the system $B$ undisturbed ($\hat E_A(i)\otimes \hat 1_B$) and vice versa, one measures the system $B$ and keeps the system $A$ undisturbed ($\hat 1_A\otimes\hat E_B(i)$). Then one takes the squared difference of these alternatives summing it over the measurement basis $\{\hat E(i)\}$. If the mean value of (\ref{divop}) is equal to zero, then we can say the systems being in the joint state $\hat\rho_{AB}$ are identical to each other with respect to the given measurement $\{\hat E(i)\}$. Otherwise, we can use the mean value of (\ref{divop}) to quantify the difference between the systems $A$ and $B$.

As the ultimate case of measurement involving all possible outcomes, let us consider an isotropic continuous POVM $\hat  1=\oint\hat E(\phi)d\nu$ which equally takes into account all states of the given Hilbert space. Here the elements of POVM are $\hat E(\phi)=\ket{\phi}\bra{\phi}$ and $d\nu$ is the differential volume of the Hilbert space. Then the divergence operator (\ref{divop}) turns to \cite{Sych:06}
\begin{equation}\label{divopcon}
\hat C_{AB}=\frac{1}{2}\oint\left(\hat E_A(\phi)\otimes \hat 1_B-\hat 1_A\otimes\hat E_B(\phi)\right)^2d\nu.
\end{equation}
In order to calculate (\ref{divopcon}), we need to specify the exact representation of a vector $\ket{\phi}$ in the $D$-dimensional Hilbert space and find the differential volume $d\nu$ of the parameter space corresponding to this representation.

To characterize a quantum state in the $D$-dimensional Hilbert space we need to fix $2D-2$ real parameters: $2D$ parameters for an arbitrary vector minus 2 parameters (the fixed norm and the overall phase). These parameters can be written as the generalized polar and azimuthal angles
\begin{equation}\label{xiD}
\xi=\left(\theta_1,\theta_2,\ldots,\theta_{D-1},\varphi_1,\varphi_2,\ldots,\varphi_{D-1}\right).
\end{equation} 
For any $D>2$ the natural representation of a state vector $\ket{\varphi}$ in the $D$-dimensional Hilbert space is
\begin{equation}\label{phiD}
\displaystyle\ket{\phi}=\left(\cos\theta_1,\ e^{i\varphi_1}\sin\theta_1\cos\theta_2 ,\ldots,\ e^{i\varphi_{D-2}}\prod\limits_{k=1}^{D-2}\sin\theta_k\cos\theta_{D-1},\ e^{i\varphi_{D-1}}\prod\limits_{k=1}^{D-1}\sin\theta_k\right),
\end{equation}
which is similar to the Bloch vector representation for the case $D=2$ \cite{Nemoto:00}. 
The differential volume in the $D$-dimensional Hilbert space is 
\begin{equation}\label{nuD}
d\nu=\left\|\frac{\partial x}{\partial \xi}\right\|d\xi=\frac{D!}{\pi^{D-1}}\prod\limits_{k=1}^{D-1}\cos{\theta_k}\sin^{2k-1}{\theta_k}d\xi,
\end{equation}
where $\xi$ is a vector of parameters (\ref{xiD}) and $x$ is a Cartesian projection of the vector (\ref{phiD}). The differential volume $d\nu$ is normalized to the dimension of the Hilbert space $D$:  $\oint d\nu=D$, where we integrate over all $\theta_k\in[0,\pi/2)$ and $\varphi_k\in[0,2 \pi)$.

After the direct calculations of the integral (\ref{divopcon}) we have a simple expression:
\begin{equation}\label{decdivop}
\hat C_{AB}={\rm\bf P}^{+}+\frac{D-1}{D+1}{\rm\bf P}^{-},
\end{equation}
where ${\rm\bf P}^+$ and ${\rm\bf P}^-$ are projectors onto symmetric and antisymmetric subspaces of the joint Hilbert space $H_{A}\otimes H_{B}$. 
In the case when the dimension of the Hilbert space of one subsystem is equal to a power of $2$, these projectors can be represented as a sum over all $\displaystyle D(D+1)/2$ maximally entangled symmetric states and over all $\displaystyle D(D-1)/2$ maximally entangled antisymmetric states, which form a basis in the joint Hilbert space $H_{A}\otimes H_{B}$ \cite{Sych:09}.

From the explicit decomposition (\ref{decdivop}) we can see that the divergence operator (\ref{divop}) is strictly positive: 
\begin{equation}\label{Climits}
1\geq\hat C_{AB}\geq\frac{D-1}{D+1}.
\end{equation}
Symmetric joint states minimize the difference between two systems, whereas the antisymmetric states maximize it. This can be easily understood in the two--dimensional case. Two qubits in the joint antisymmetric singlet Bell state exhibit anticorrelations which are exactly opposite to correlations. Therefore two subsystems in the overall singlet joint state look maximally different with respect to the identical measurements. To minimize this difference, systems must be in a symmetric joint state. As far as the basis of the symmetric subspace of the joint Hilbert space $H_{A}\otimes H_{B}$ can be composed of any symmetric states (not necessarily maximally entangled), all joint symmetric states have the same minimum mean value of divergence. A particular case of symmetric states is a product state composed of two identical states $\ket{\varphi}_A\ket{\varphi}_B$. Due to the nonzero value of divergence, two systems although in the same identical state are not identical with respect to the measurement results. One can say, that any two quantum systems are always different, even though they are in the same quantum state.

Let us note, that the quantum uniqueness can be interpreted in two ways. One can say that it is a limitation imposed by quantum measurements, namely the impossibility of controlling their particular outcome at a given time. We would like to mention, however, that the uniqueness issue is not restricted to the isolated measurement process itself. Any arbitrary dynamics of the system can be described as a process of transformations possibly involving other physical systems, but finally one needs to see the classical results, i.e. to measure the final state of the system. In all the situations where this measurement involves a non orthogonal set of states, explicitly or implicitly, the uniqueness issue applies. Thus it should not be regarded as an imperfection or a limitation of quantum measurements, but must be considered as the essential distinctive property of the quantum world, similarly to ``no--cloning''.

Quantum uniqueness demonstrates an intrinsic difference between any two quantum systems, which gives rise to a number of interesting applications. We note one of them~--- the generation of objectively true random numbers. There are many definitions of randomness, each of them highlighting a particular aspect of this complex notion. We call a system as {\em objectively random} if there is no other system that would perfectly predict or reproduce its objective properties, e.g. the measurement results of the system. In other words, the perfect copy of objectively random system does not physically exist. Indeed, if one had a perfect copy of the system, then by looking at the copy one could predict the behavior of the system, particularly the result of a measurement. To be more specific, let us consider a random number generator as a particular system of interest. If one has another system which is a perfect copy, then a sequence of numbers at the output of the random number generator may look chaotic and truly random in a statistical sense. But in fact it will be perfectly correlated with the other system (its copy), therefore it is not objectively random. Otherwise, if there can be no such perfect copy, then the sequence of numbers at the output of random number generator is objectively random: it cannot be perfectly predicted by any other system and it is asymptotically unique. As a consequence, it lacks any order and is also truly random in a mathematical sense.

Existence of objective randomness is naturally inspired by quantum uniqueness, but does not immediately follow from it. Uniqueness in the above discussed sense does not mean unpredictability of quantum measurements. Let us consider, for example, a simple measurement: detection of a linearly $H/V$ polarized photon which can be either reflected or transmitted by a $\pm 45^\circ$ oriented polarization beam splitter. Appearance of these two alternatives (transmission or reflection) is widely believed to be truly random, according to the Copenhagen interpretation of quantum mechanics. However, if one prepares an entangled pair of photons in the joint polarization singlet state $\ket{\Psi^-}=(\ket{H}\ket{V}-\ket{V}\ket{H})/\sqrt{2}$ and sends this pair to the same polarization beam splitter (or two different beam splitters with identical orientation), then the measurement outcomes become strictly correlated. The probability of an event when both photons are either reflected or transmitted (let us denote it as $\ket{x}$) is zero: $P_{AB}(x,x)={\rm Tr}\left[(\ket{x}_A\bra{x}_A\otimes\ket{x}_B\bra{x}_B )\ket{\Psi^-}_{AB}\bra{\Psi^-}_{AB}\right]={\rm Tr}[\ket{x}_A\ket{x}_B\bra{\Psi^-}_{AB}(\braket{x}{H}_A\braket{x}{V}_B-\braket{x}{V}_A\braket{x}{H}_B)]/\sqrt{2}=0.$ Hence one photon will be reflected, and another one will be transmitted. Thus the measurement result of the first photon, being truly random, can be used to predict another truly random result of the same measurement of the second photon. Here we can see a clear difference between true randomness, which mostly refers to absence of any reason for a given outcome, and objective randomness, which means principal physical unpredictability of the outcome and therefore its uniqueness.

Intrinsic true randomness of quantum measurements has been widely studied since the famous discussion between Einstein and Bohr \cite{EPR:35,Bohr:35}. Despite the rather abstract character of this area, it has one particularly useful and natural application~--- generation of true random numbers. The problem of potential predictability of random number generators based on quantum measurements was pointed out recently \cite{Fiorentino:07,Svozil:09,Pironio:09}. In our terms, objective randomness implies true randomness, but not vice versa. In certain applications (e.g. quantum key distribution), predictability of random numbers can be a serious issue. To avoid the problem, it was suggested to use pure quantum states \cite{Fiorentino:07}, to make non orthogonal measurements in three-- and higher--dimensional Hilbert spaces \cite{Svozil:09} or to rely on violation of Bell's inequality \cite{Pironio:09}.

In our approach, to get unpredictable (i.e. objectively random) measurement outcomes one has to compare the two {\em most correlated} measurements instead of two {\em identical} measurements in the operator (\ref{divopcon}) which leads to the modified operator 
\begin{equation}\label{divopcon2}
\hat C_{AB}=\frac{1}{2}\oint\left(\hat E_A(\phi)\otimes \hat 1_B-\hat 1_A\otimes\hat E_B\left(f_{\rho}(\phi)\right)\right)^2d\nu,
\end{equation}
where $f_{\rho}(\phi)$ is a function which depends on $\hat\rho_{AB}$ and the state $\ket\phi$ in order to minimize the mean value of (\ref{divopcon2}). For maximally entangled states one can always find such a function that the mean value of (\ref{divopcon2}) is equal to zero. For example, in the simplest case of antisymmetric singlet Bell state the function $f(\phi)$ is just inversion ($f(\phi)=\tilde\phi$), as we have seen above. On the other hand, for joint product states $\hat\rho_{AB}=\hat\rho_{A}\otimes\hat\rho_{B}$ any two local quantum measurements are always uncorrelated because the joint probability is in a product form:
\begin{equation}\label{PabPaPb}
P_{AB}(\varphi,f(\varphi))={\rm Tr}[(\hat E_{A}(\varphi)\otimes\hat E_{B}(f(\varphi)))(\hat\rho_{A}\otimes\hat\rho_{B})]={\rm Tr}[\hat E_{A}(\varphi)\hat\rho_{A}]{\rm Tr}[\hat E_{B}(f(\varphi))\hat\rho_{B}]=P_{A}(\varphi)P_{B}(f(\varphi)).
\end{equation}
Here we can notice that a joint state of one pure partial state and any other partial state is always in a product form, thus it cannot be entangled. Moreover, the measurement results of a pure state are completely uncorrelated to the measurement results of any other system, therefore they are objectively unpredictable. Here we conclude that by measuring a pure state one can guarantee objective randomness of the measurement outcomes. We don't need to specify a particular kind of measurement. In principle, any type is suitable under the assumption that the measured state is pure. In practice, it is convenient to combine randomness generation with purity check by non orthogonal measurement. It is interesting to note, that at this point we come to almost the same conclusion as in \cite{Fiorentino:07}. 

To summarize, we have pointed out a special property of quantum states~--- the uniqueness of quantum systems with respect to their possible measurement outcomes. The formal description of uniqueness is done by calculating the divergence operator (\ref{divopcon}). Its explicit decomposition (\ref{decdivop}) shows that any two systems cannot deliver universally identical measurement results, involving all possible outcomes, or all states of the given Hilbert space. We also conjecture, that the observed phenomenon is valid with respect to any non orthogonal measurement, not necessarily the isotropic continuous one. Finally, we have found a tight connection of quantum uniqueness with objective randomness of quantum measurements and revealed a simple way for generation of objective randomness by measuring a pure quantum state.

D.S. acknowledges Alexander von Humboldt Foundation for a stipend.


\end{document}